\documentclass[12pt]{article} 
\begin{document}
\begin{center}
\large{\textbf{Quantum Gravity Effects in Geodesic Motion \\ and Predictions 
of 
Equivalence Principle Violation}}\\
\end{center}
\begin{center}
Subir Ghosh\\
Physics and Applied Mathematics Unit, Indian Statistical
Institute\\
203 B. T. Road, Kolkata 700108, India \\
\end{center}
\vspace{0.2cm}

\textbf{Abstract:} We show that the Equivalence Principle (EP) is violated by
Quantum Gravity (QG) effects. The predicted violations are
compared to experimental observations for Gravitational Redshift, Law of
Reciprocal Action and Universality of Free Fall. This allows us to derive
explicit  bounds  for
 $\beta$ - the QG scale. 

In our
approach, there appears a deviation
in the geodesic motion of a particle. This deviation is induced by a
non-commutative spacetime, consistent with
a Generalized Uncertainty
Principle (GUP). Gup admits the presence of a minimum length scale, that is
advocated by QG
theories. Remarkably, the GUP 
induced corrections are quite robust since the bound on $\beta$ obtained by us,
{\it{in General Relativity scenario in an essentially classical setting}} of
modified geodesic motion,
is closely comparable to similar bounds  in recent literature \cite{vag}. The
latter are computed in
purely {\it{quantum}}  physics domain in {\it{flat}} spacetime.

\vskip .5cm
{\bf {Introduction}}: Modern theory of Gravitation is essentially Einstein's
theory of General Relativity which is based on a key 
concept: the Equivalence Principle (EP) \cite{ein}. It states that there is no
way to differentiate between uniformly
accelerated reference frame and  gravitation using local measurements.
Exploiting EP one can derive the geodesic motion of a particle in presence of
gravity from the rectilinear
motion of the same particle in freely falling coordinate system simply by
replacing the flat Minkowski metric in the latter to a general metric
$g_{\mu\nu}$. An alternative
manifestation of EP, especially convenient in low energy (Newtonian) domain, was
suggested by Bondi \cite{bon}, where
the idea of active and passive masses of a particle was introduced and their
inequality, (if observed experimentally), would signal EP violation. Most
of the terrestrial experiments \cite{lam,ash,bar}
showing validity of EP are based on the latter scenario. However, the Newtonian
potential approach \cite{bon} is derived as a low energy limit of the basic
geodesic motion so, quite
obviously, any violation in EP in Newtonian physics has to be present as a
deviation in the geodesic motion (which is more
fundamental). In this Letter we use this top-down approach and show that QG
effects induce a deviation
in the geodesic motion that can have low energy experimental consequences
related to EP violation as in
\cite{lam,ash,bar}.

The fact that taking account of quantum phenomena can affect EP has been
established in \cite{doug} by showing
that a Unruh-DeWitt detector can distinguish between gravitational field and an
"equivalent" accelerating
reference frame. It was argued in \cite{doug} that the incompatibility between
quantum phenomena and EP appears because
the former is  inherently non-local (Bell entanglement, uses  non-local plane
wave modes in field theory), 
whereas EP assumes local measurements. {\it{This argument strongly favours the
possibility of EP violation by
QG effects since QG theories are all the more non-local as they unanimously
advocate an absolute short distance
scale- the Planck scale}} (see also \cite{am1}). Other
forms of this clash has been suggested in \cite{ot} in the context of neutrino
oscillations.

The major obstacle in experimantal verification of QG effects is their smallness
since
the predicted corrections are scaled by Planck scale of energy. Some positive
indications have appeared recently \cite{am} in testing QG induced modified
energy-momentum
dispersion relations in  cold-atom-recoil experiments. On the other hand, as
pointed out in \cite{vag},
GUP effects are quite universal in nature and its  predictions
 can provide large upper bounds for $\beta_0$, (the GUP parameter
$\beta =\beta_0/M^2_{Planck}$),
consistent with present day experimental observations and suggest the existence
of a new scale between electroweak
scale and Planck scale. This is very relevant to our work since  we have also
provided upper bounds similar to \cite{vag}) for $\beta_0$. It is very
significant
that whereas, \cite{vag}
{\it{ deals with purely flat spacetime quantum phenomena}}, such as  Lamb
Shift, 
Landau levels and
the
tunnelling current in a Scanning Tunnelling Microscope, {\it{we work in General
Relativity in an
essentially classical framework}} of (QG modified) action and variational
(geodesic) equations of motion and still come up with comparable and improved
predictions w.r.t. \cite{vag}. {\it{This
underlines robustness of the QG effects induced by GUP.}}

QG theories, (such as String Theory or Loop Quantum Gravity),  predict a
Generalized Uncertainty
Principle (GUP)  with a minimum length scale \cite{str}. This is compatible with
a
Non-Commutative (NC) spacetime as derived in  \cite{kem} in Minkowski spacetime.
 {\it{This is the key point in our framework: QG effects enter because we are
exploiting a NC algebra that induces a minimum length scale through GUP.}}
We generalize the above NC algebra to curved
spacetime and 
 derive, for the first time, from
first principles in a Hamiltonian framework, the modified (QG or GUP corrected)
geodesic equation satisfied by a point particle. Working in weak linearized
gravity and Newtonian low energy limit
we show that in the leading order of NC parameter $\beta $
the modified geodesic equation depends on the particle mass. {\it{Clearly this
is a
violation of the EP.}} Previous attempts to show GUP effects on EP  \cite{ali}
were restricted to
Newtonian physics without considering
the geodesic deviation. The latter was treated in \cite{hari} in a
$\kappa$-Minkowski spacetime that is different from ours. As we argued above it
seems natural that QG effects can violate
EP, it is paradoxical from String Theory perspective since it advocates 
 GUP  \cite{str}, (due to the short
distance scale), but at the same time
String Theory in the low energy limit should yield Einstein's General Relativity
(that
is based on EP).\\  
{\bf{QG/GUP inspired model in curved spacetime}}: 
  So far the GUP oriented studies have been mostly
kinematical  but to analyze the dynamics it is essential to have a relativistic
point particle
Lagrangian/Hamiltonian framework as in \cite{sp,spint}, which we have extended
here from flat to curved spacetime. 
There is some  ambiguity involved in the explicit form of the generalized
particle model because there can be many inequivalent extensions all of which
induce GUP-type  phase
space and also reduce to the canonical particle model for $\beta =0$. Only
experimental results can distinguish one model from the other. We have chosen
the simplest model, with commuting coordinates 
even in presence of gravity, but  retaining the GUP induced minimal length
feature.

The free GUP particle model in flat spacetime \cite{sp}
 \begin{equation} L=-A\eta_{\mu\nu}x^\mu\dot p^\nu
 +\beta (xp)(p\dot p)
\label{free} \end{equation}
with $A=1-\beta \frac{p^2}{2},~ p^2=(p^0)^2-(\vec p)^2,~(ab)=\eta_{\mu\nu}a^\mu
b^\nu $ satisfies the NC
algebra \cite{kem},
\begin{equation} \{x^\mu,p^\nu\}=-\left[\frac{g^{\mu\nu}}{\left(1-\frac{\beta
p^2}{2}\right)}+\frac{\beta p^\mu p^\nu}{\left(1-\frac{3\beta
p^2}{2}\right)\left(1-\frac{\beta p^2}{2}\right)}\right], ~
 \{x^\mu,x^\nu\}=\{p^\mu,p^\nu\}=0. \label{db2} \end{equation}
It is easy to check that this algebra is Lorentz covariant by noting  that for
infinitesimal Lorentz transformations, $x^{'\mu} =
x^\mu +\delta \omega^\mu_\nu x^\nu ,~p^{'\mu }=
p^\mu +\delta \omega^\mu_\nu p^\nu ;~\delta \omega_{\mu_\nu}=-\delta
\omega_{\nu_\mu}$ the full algebra remains
form invariant, with $x^\mu,p^\mu$ replaced by $x^{'\mu},p^{'\mu }$ and
$p^2=p^{'2}$:
$$\{x^{'\mu},p^{'\nu}\}=-\left[\frac{g^{\mu\nu}}{\left(1-\frac{\beta
p{'^2}}{2}\right)}+\frac{\beta p{'^\mu} p^{'\nu}}{\left(1-\frac{3\beta
p{'^2}}{2}\right)\left(1-\frac{\beta p{'^2}}{2}\right)}\right], ~
 \{x{'^\mu},x^{'\nu}\}=\{p^{'\mu},p^{'\nu}\}=0. \label{db2}.
$$
Furthermore, to $O(\beta)$ that we will adhere to later, the (deformed) Lorentz
generators $L^{\mu\nu}=(1-\frac{\beta p^2}{2})
(x^\mu p^\nu -x^\nu p^\mu )$ yield $\delta x^\mu =\frac{\delta \omega
_{\alpha\beta}}{2}\{L^{\alpha\beta},x^\mu \},~
\delta p^\mu =\frac{\delta \omega _{\alpha\beta}}{2}\{L^{\alpha\beta},p^\mu \}$
and satisfy the undeformed Lorentz algebra and provide an invariant mass-shell
condition $p^2=m^2$. Similar form of Lorentz covariantization of the original
non-relativistic 
GUP algebra \cite{kem} has already appeared in \cite{que}. In \cite{que} it is
pointed out that 
indeed the covariant algebra leads to a minimum length GUP, similar to
\cite{kem},  although it
does not reduce to  \cite{kem}
in non-relativistic limit.

In the presence of gravity, this is generalized to ($\eta_{\mu\nu}\rightarrow
g_{\mu\nu}$),
\begin{equation} L=-Ag_{\mu\nu}x^\mu\dot p^\nu
-(\partial_\lambda g_{\mu\nu})p^\mu x^\nu\dot x^\lambda +\beta (xp)(p\dot p).
\label{l} \end{equation}
This is a first order system with constraints and the Dirac Hamiltonian scheme
\cite{dir}
is used to obtain the Dirac Brackets to first order in $\beta$,
\begin{equation} \{x^\mu ,x^\nu\}= 0 ~;~\{p^\mu ,p^\nu\}=Q^{\mu\nu}+\beta
(H^{\mu\nu}-Q^{\mu\lambda}M_{\lambda\sigma}g^{\sigma\nu}-
g^{\mu\lambda}M_{\sigma\lambda}Q^{\sigma\nu}),$$$$\{x^\mu
,p^\nu\}=A^{-1}g^{\mu\nu}+\beta (cp^\mu p^\nu
-g^{\mu\lambda}M_{\lambda\sigma}g^{\sigma\nu}) \label{db1} \end{equation}
where the abbreviations are,
\begin{equation} c=\frac{\beta}{A(A-\beta p^2)},~
Q^{\alpha\lambda}=g^{\alpha\mu}g^{\lambda\nu}(\partial_\mu
g_{\nu\sigma}-\partial_\nu g_{\mu\sigma} )p^\sigma~,$$$$ \beta
H^{\alpha\lambda}=\beta [(\frac{p^2}{2}g^{\alpha\mu}+p^\alpha p^\mu
)g^{\nu\lambda} +(\frac{p^2}{2}g^{\nu\lambda}+p^\nu p^\lambda
)g^{\alpha\mu}](\partial_\mu g_{\nu\sigma}-\partial_\nu
g_{\mu\sigma})p^\sigma,$$$$
M_{\mu\nu}=-(\frac{1}{2}g_{\alpha\beta}\partial_\nu g_{\mu\lambda}+
\frac{1}{2}g_{\mu\lambda}\partial_\nu g_{\alpha\beta} +g_{\alpha\mu}\partial_\nu
g_{\lambda\beta}
+g_{\alpha\lambda}\partial_\nu g_{\mu\beta})p^\alpha p^\beta x^\lambda .
\label{xx}
\end{equation}
This non-canonical algebra appears as Dirac Brackets (with details  in 
Appendix.) For $g_{\mu\nu}\rightarrow
\eta_{\mu\nu}$
the flat space  GUP model (\ref{free},\ref{db2}) is recovered. The curvature
corrected GUP
algebra
is  one of our major results, similar to the $U(1)$ gauge interaction extension
discussed in
\cite{spint}. Note that this algebra will obviously lead to  GUP like minimum
length uncertainty relation
with (derivative of metric) corrections. Hence the dynamics, derived below, is
considered to be Quantum Gravity corrected.

The Hamiltonian equations of motion are obtained from,
\begin{equation}
\dot x^\mu =\{x^\mu ,H\}=g_{\nu\lambda }p^\lambda\{x^\mu ,p^\nu\} ~;~\dot p^\mu
=\frac{1}{2}p^\nu p^\lambda \{p^\mu ,g_{\nu\lambda }\} +g_{\nu\lambda
}p^\lambda\{p^\mu ,p^\nu\},
\label{dot}
\end{equation}
with the Hamiltonian constraint given by,
\begin{equation}
H=\frac{1}{2}(g_{\mu\nu}p^\mu p^\nu -m^2).
\label{h}
\end{equation}
So far we have not done any approximation regarding $g_{\mu\nu}$. Now we
linearize: $
g_{\mu\nu }=\eta_{\mu\nu }+h_{\mu\nu }+O(h^2).$
This yields slightly simplified forms of the equations of motion:
\begin{equation}
\dot p^\mu =\eta^{\mu\nu} (\frac{1}{2}\partial_\nu h_{\rho\sigma
}-\partial_\sigma h_{\nu\rho })p^\rho p^\sigma
+\beta (2m^2\eta^{\mu\nu}+p^\mu p^\nu )(\frac{1}{2}\partial_\nu h_{\rho\sigma
}-\partial_\sigma h_{\nu\rho })p^\rho p^\sigma ,
\label{dotp}
\end{equation}
\begin{equation}
\dot x^\mu =p^\mu +\beta (\eta _{\nu\rho}+h_{\nu\rho}) H^{\mu\nu }p^\rho .
\label{dotx}
\end{equation}
To $O(\beta )$ we  invert the above equations to
get a modified geodesic equation,
\begin{equation}
\ddot x^\mu =[(1+\frac{5}{2}\beta m^2)\eta^{\mu\nu}\dot x^ \rho \dot x^\sigma
-\beta \frac{m^2}{2}\dot x^\mu
(\dot x^\sigma \eta ^{\rho\nu }+\dot x^\rho \eta^{\sigma\nu})]
(\frac{1}{2}\partial_\nu h_{\rho\sigma }-\partial_\sigma h_{\nu\rho }).
\label{ddotx}
\end{equation}
For $\beta =0$ the geodesic equation is reproduced. 
In the above we have used the constraint $p^2=m^2$. Presence of $m$ signals EP
violation. {\it{This is our most important result}}.\\
{\bf{Predictions of EP Violation}}: 
We wish to predict terrestrially observable effects of EP violation in our
model and so consider low energy Newtonian limit. Renaming the parameter $\beta
m^2/2 =\beta_m$ from (\ref{ddotx}) we find,
\begin{equation}
\frac{d^2t}{d\tau
^2}=\beta_m(\frac{dt}{d\tau})^2\frac{dx^i}{dt}\eta^{\mu\nu}\partial_i
h_{\mu\nu} \approx 0,
\label{x0}
\end{equation}
\begin{equation}
\frac{d^2x^i}{dt^2}=\frac{1}{2}(1+5\beta_m)\partial^ih_{00}-\beta_m
\frac{dx^i}{dt}\frac{dx^j}{dt}(\eta^{00}\partial_jh_{00}+\eta^{kl}\partial_jh_{
kl})
\approx \frac{1}{2}(1+5\beta_m)\partial^ih_{00}.
\label{d2x}
\end{equation}
In the approximate equalities velocity terms are dropped. We observe that the GUP induced deviation scales quadratically with the mass of the test paricle whereas theoretical results in literature for Weak Equivalence Principle violaion generally depend linearly on the mass.\\
{\it{Gravitational Redshift}}: In the conventional case ($\beta_m =0$)  from
Newton's equation and gravitational potential at a distance
$r$ from a mass $M$, $(d^2\mathbf x)/(dt^2)=-\nabla \phi ;~\phi =-GM/r,$
one identifies $h_{00}=-2\phi ~ \rightarrow g_{00}=-(1+2\phi)$ (see eg.
\cite{wein}).
In the present case we have $(1+5\beta _m)h_{00}=-2\phi $ so that $h_{00}\approx
-2(1-5\beta _m)\phi$ leading to
$g_{00}=-(1+2(1-5\beta _m)\phi )$.

In order to experimentally measure Gravitational Redshift effect \cite{wein} one
needs two observation points, say $x_1,x_2$ and
consider a given atomic transition. The ratio of frequencies $\nu_2$ - light
coming from $x_2$ to $x_1$, and $\nu_1$, both observed at $x_1$, is
\begin{equation}
 \frac{\nu(x_2)}{\nu(x_1)}=\left(\frac{g_{00}(x_2)}{g_{00}(x_1)}\right
)^{\frac{1}{2}}=\left(\frac{
1+2(1-5\beta_m)\phi(x_2)}{1+2(1-5\beta_m)\phi(x_1)}\right)^{\frac{1}{2}} \approx
1+(1-5\beta_m)(\phi(x_2)-\phi(x_1)) ,
\label{exp}
\end{equation}
where the above expression is linearized in the last step. Hence for two clocks
$A$ and $B$ \cite{lam}, with $(\beta_m)_A=(\beta m_A^2)/2$,.., we
will have
\begin{equation}
 \frac{\nu_A(x_2)}{\nu_A(x_1)} \approx 1+(1-5(\beta_m)_A)(\phi(x_2)-\phi(x_1));~
\frac{\nu_B(x_2)}{\nu_B(x_1)} \approx 1+(1-5(\beta_m)_B)(\phi(x_2)-\phi(x_1)).
\label{exp1}
\end{equation}
Combining the above expressions we obtain the all important result \cite{lam},
\begin{equation}
\left(\frac{\nu_A(x_2)}{\nu_B(x_2)}\right)\approx
\{1-5((\beta_m)_A-(\beta_m)_B)\}(\phi(x_2)-\phi(x_1))
\left(\frac{\nu _ A(x_1)}{\nu _ B(x_1)}\right).
\label{02}
\end{equation}
A mismatch of the frequency ratios will signal a violation of the EP. The
best present day observational result
is $\mid \alpha_{Hg}-\alpha_{Cs} \mid \leq 5.10^{-6}$ \cite{ash} where
$\alpha_{Hg},~\alpha_{Cs}$ stand for clock-dependent parameters for Mercury and
Cesium (for details see \cite{lam,ash}). In our case
$\alpha_{Hg}\equiv 5\beta m^2_{Hg},~\alpha_{Cs}\equiv 5\beta m^2_{Cs}$.
Conventionally one considers $\beta =\beta_0/M^2_{Planck}$  \cite{vag} with
$\beta_0\approx
1$, in which case the mismatch will be
$\approx (m^2_{Hg}- m^2_{Cs})/M^2_{Planck}\approx 10^{-34}$ {\footnote{I thank
Prof. Douglas Singleton for pointing out an error in the
numerical estimate in an earlier version of the paper.}}. Indeed this signal
is very small. Another interpretation \cite{vag} is
to consider an upper bound for $\beta_0$: $\beta_0\leq
(10^{-9}/10^{-25})^2.10^{-6}\approx 10^{28}$. This is below
the upper bound of $\beta_0\leq 10^{34}$ compatible with the electroweak
scale and same as the bounds suggested in \cite{vag} from Lamb shift
and Landau level measurements, but weaker than $\beta_0\leq 10^{21}$, again
derived
from Scanning Tunneling Microscope current
measurement \cite{vag}.\\
{\it{Law of Reciprocal Action}}: The notion of distinct masses was introduced by
Bondi where the (Newtonian) gravitational force law between two masses $A,B$  is
generalized to
\begin{equation}
  m_{Ai}\ddot\mathbf x_A=Gm_{Ap}m_{Ba}\frac{\mathbf x_B-\mathbf x_A}{\mid
\mathbf x_B-\mathbf x_A\mid^3},~
m_{Bi}\ddot \mathbf x_B=Gm_{Bp}m_{Aa}\frac{\mathbf x_A-\mathbf x_B}{\mid \mathbf
x_B-\mathbf x_A\mid^3}.
\label{new}
\end{equation}
In the above force law  for  $A$ \cite{lam} $m_{Ai}$ is the {\it{inertial}}
mass, $m_{Ap}$ is
the {\it{passive}} mass and $m_{Aa}$ is
the {\it{active}} mass as they appear in 
$m_{Ai}\ddot \mathbf x=m_{Ap}\nabla U(\mathbf x);~\nabla ^2U(\mathbf x)=4\pi
m_{Aa}\delta(\mathbf x).$
The motion of the center of mass coordinate
$\mathbf X=(m_{Ai}\mathbf x_A+m_{Bi}\mathbf x_B)/(m_{Ai}+m_{Bi})$ is given by
\begin{equation}
\ddot\mathbf X=G\frac{m_{Ap}m_{Bp}}{m_{Ai}+m_{Bi}}C_{BA}\frac{\mathbf
x_B-\mathbf x_A}{\mid \mathbf
x_B-\mathbf x_A \mid^3},~C_{BA}=\frac{m_{Ba}}{m_{Bp}}-\frac{m_{Aa}}{m_{Ap}}.
\label{05}
\end{equation}
For $C_{BA}\neq 0$ the center of mass will possess a self-acceleration
\cite{lam}.  In our formlation the potential and hence the active mass gets
modified so that
\begin{equation}
 C_{BA}=\frac{m_{Ba}}{m_{Bp}}-\frac{m_{Aa}}{m_{Ap}}=\frac{(1-5(\beta_m)_B)m_{Bi}
}{m_{Bi}}-\frac{(1-5(\beta_m)_A)m_{Ai}}{m_{Ai}}$$$$
=5((\beta_m)_B-(\beta_m)_A)=5\beta_0\frac{m_B^2-m_A^2}{M_{Planck}^2}.
\label{ex1}
\end{equation}
Observation
of no self-accleration of the moon
by Lunar Laser Ranging provides a bound $\mid C_{Al-Fe}\mid\leq 7.10^{-13}$
\cite{bar,lam}. This provides a considerably tighter bound $\beta_0\leq 10^{21}$
than the one provided by Gravitational Redshift (see above)
and is of the same order as earlier bounds \cite{vag}. {\it{This is our most
important
prediction}}.\\
{\it{Universality of Free Fall}}: According to General Relativity the neutral
free particles follow the geodesic and hence 
the motion is independent of the nature of the neutral particle. Its' validity
is tested by experimentally 
measuring the Eotovos parameter $\eta =\frac{g_A-g_B}{\frac{1}{2}(g_A+g_B)}$
where $g_A,g_B$ are accelerations of two 
particles $A$ and $B$ in the ``same'' gravitational field. A non-zero $\eta$
signals violation of Universality of Free Fall. But in the present case the
active mass gets different corrections
for $A$ and $B$ and in turn the gravitational field perceived by them is not the
same. In the field of $M$ the acceleration of $A$ is $g_A=(1-5(\beta_m)_A)g$
(and similarly for $B$). Thus we find
\begin{equation}
 \eta
=\frac{(1-5(\beta_m)_A)-(1-5(\beta_m)_B)}{\frac{1}{2}
(1-5(\beta_m)_A)+(1-5(\beta_m)_A)}\approx 5\beta_0 (m^2_B-m^2_A)/M^2_{Planck}.
\label{uff}
\end{equation}
Tosion pendulum results provide $\eta \leq 2.10^{-13}$ \cite{lam} (for the AL-FE
pair) yielding once
again $\beta_0\leq 10^{21}$.
Note that the results will not
hold for macroscopic bodies due to the restriction $\beta_m<<1$. \\
{\bf{Conclusion}}: We have shown that, at least in principle, Quantum Gravity
effects will lead to violation of Equivalence Principle because even in the low
energy and weak gravity limit,
QG/GUP modifies gravitational potential obtained from deviated geodesic
equation. The correction
depends upon
the test particle energy/mass signalling  a violation of Equivalence
Principle. The experimental signature scales with the square of the mass of the test particle mass.
 Results are predicted for the
violation in the contexts of Gravitational Red Shift, Law of Reciprocal
Action and Universality of Free Fall. Comparison with
experimental results predict explicit bounds for the GUP parameter. The GUP
corrections are quite robust since our results, $\beta_0\leq 10^{21}$ in General
Relativity (in a classical setting) agree with previous predictions \cite{vag}
(obtained
in flat spacetime purely quantum
 phenomena), which is indeed remarkable. Our analysis and results can have
serious consequences in the context of String Theory since, on the one hand
String Theory requires a
modification in the Heisenberg Uncertainty Principle, and advocates some form of
Generalized Uncertainty Principle, (with a short distance scale), but on the
other hand String Theory is also expected to
reduce to Einstein General Relativity,  which essentially rests on Equivalence
Principle. Our analysis shows
that particle dynamics, taking in to account the Generalized Uncertainty
Principle is not compatible with Equivalence Principle.

 Lastly we stress that
Quantum Gravity effects appear most
natural in geodesic deviations, as demonstrated here, since they are directly
linked to metric derivatives and would appear only 
in an ad hoc way in flat space computations.

\vskip .05cm
{\it{Appendix}}: We briefly discuss steps leading to the Dirac Brackets. In the
presence of a set of Second Class Constraints $\psi_\mu$,
with non-singular  constraint algebra matrix $\{\psi_\mu, \psi_\nu
\}$, the Dirac bracket between two generic
variables $A$ and $B$ is defined as
\begin{equation}
 \{A,B\}_{DB}=\{A,B \}-\{A, \psi_\mu \}\{\psi_\mu, \psi_\nu \}^{-1}\{\psi_\nu, B
\}.
\label{db}
\end{equation}
We have dropped the subscript $\{,\}_{DB}$ throughout.

From the Lagrangian the  conjugate momenta and constraints
$\phi^1_\mu,\phi^1_\nu$ are obtained:
\begin{equation} \pi^x_\mu = \frac{\partial
L}{\partial \dot x^\mu}=-\partial _\mu g_{\alpha\beta}p^\alpha x^\beta~;~
\pi^p_\mu = \frac{\partial L}{\partial \dot p^\mu}=-Ag_{\mu\nu}x^\nu+\beta
(xp)p_\mu,
\label{p} \end{equation}
\begin{equation} \phi^1_\mu \equiv \pi^p_\mu +Ag_{\mu\nu}x^\nu -\beta (xp)p_\mu
\approx
0~;~\phi^2_\mu \equiv \pi^x_\mu +\partial _\mu g_{\alpha\beta}p^\alpha
x^\beta \approx 0 .\label{c} \end{equation}
 The following algebra shows that the constraints are Second Class,
\begin{equation}
\{\phi^1_\mu,\phi^1_\nu\}=0,~\{\phi^2_\mu,\phi^2_\nu\}=(\partial_\mu
g_{\nu\alpha}-\partial_\nu g_{\mu\alpha})p^\alpha,~
\{\phi^1_\mu,\phi^2_\nu\}=Ag_{\mu\nu}-\beta p_\mu p_\nu +\beta
M_{\mu\nu}, \label{cc}
\end{equation}
The constraint matrix is  $\{\phi_\mu^i,\phi_\nu^j\}=A+\beta B$,
\begin{equation}
  A = \left[ {\begin{array}{cc}
 0 & (Ag_{\mu\nu}-\beta p_\mu p_\nu) \\
 -(Ag_{\mu\nu}-\beta p_\mu p_\nu)  & (\partial_\mu g_{\nu\alpha }-\partial_\nu
g_{\mu\alpha })p^\alpha \\
 \end{array} } \right]
~,~~
  B = \left[ {\begin{array}{cc}
 0 & \beta M_{\mu\nu} \\
 -\beta M_{\nu\mu} & 0\alpha \\
 \end{array} } \right] ,
\label{matr}
\end{equation}
yields the inverse $(A+\beta B)^{-1}\approx A^{-1} -\beta A^{-1}BA^{-1}$
to first order in $\beta$. The Dirac Brackets are computed from (\ref{db}).\\

\end{document}